\documentclass[twoside]{article}

\usepackage[T2A]{fontenc}
\usepackage[utf8x]{inputenc}
\usepackage[english]{babel}
\usepackage[a4,frame,center]{crop}
\usepackage{hyperref}

\usepackage{amssymb,graphicx,rotating,multirow,multicol}
\usepackage{amsmath,amsthm,amsfonts,amscd}
\usepackage[mathscr]{euscript}
\newcommand{\be}{\begin{equation}}
\newcommand{\ee}{\end{equation}}
\newcommand{\bea}{\begin{eqnarray}}
\newcommand{\eea}{\end{eqnarray}}
\newcommand{\lb}{\label}
\newcommand{\bdm}{\begin{displaymath}}
\newcommand{\edm}{\end{displaymath}}
\newcommand{\D}{{\rm d}}

\usepackage{lipsum}
\usepackage[margin=1in, textheight=210mm, textwidth=250mm]{geometry}
\usepackage{multicol}
\begin{document}
\begin{center}
  {\huge The impact of Friedmann's work on cosmology}\\
  \vskip 3mm
{\Large Claus Kiefer\footnote{University of Cologne, Faculty of
    Mathematics and Natural Sciences, Institute for Theoretical
    Physics, Cologne, Germany.} and Hermann Nicolai\footnote{Max
    Planck Institute for Gravitational Physics, Albert Einstein
    Institute, Potsdam, Germany.}}
\end{center}

\vskip 4mm

\begin{multicols}{2}

  \noindent {\bf Historical introduction.}
 The impact of Friedmann's work on cosmology can hardly be
 overestimated. By training, Friedmann was a mathematician, but one of exceptional
versatility, who made important contributions also in other fields, such
as meteorology. In the summer of 1917 and in the middle of tumultuous
events in Russia, he founded and was the first director of the
``Aeropribor'' factory in Moscow which produced tools for airplanes, and
which still exists to this day. Nevertheless, his greatest
contribution 
to science is undoubtedly contained in the two pioneering paper in
1922 and 1924 which appeared in the German
journal {\em Zeitschrift f\"ur Physik} [Friedmann~1922,
1924].\footnote{In [Friedmann~1922], the German transcription of the
  Russian name was chosen ``Friedman'', but we stick to the common practice of
  writing ``Friedmann''. For an editorial note to the English
  translation of these papers, see [Ellis~1999].}
In these papers, he demonstrated that Einstein's field equations with
a cosmological constant (called by him {\em Weltgleichungen},
i.e. world equations) do not only allow Einstein's 1917 static
solution with matter and de~Sitter's 1917 apparent static vacuum
solution, but also dynamical solutions describing an expanding or
collapsing Universe. The corresponding equations, today called
Friedmann or Friedmann--Lema\^{\i}tre equations, form the basis of
modern cosmology. In 1923, Friedmann published a book on cosmology in
which he also presents insights into his general philosophical ideas
[Friedmann 2000].\footnote{The editor of the German translation
  [Friedmann 2000]
  speculates that the title \emph{\foreignlanguage{russian}
  {Мир как пространство и время}} (Die
  Welt als Raum und Zeit) alludes to Schopenhauer's opus magnum \emph{\foreignlanguage{russian}
    {Мир как воля и представление}} (Die Welt als Wille und Vorstellung).} 

In the 1920s, Friedmann's work had little impact [Ellis 1989]. The
main question in those years was trying to find out whether there is an
observational difference between Einstein's and de~Sitter's
solution. Friedmann's papers were apparently also unkown to Georges
Lema\^{\i}tre, who in 1927 wrote another groundbreaking paper that was
little appreciated at the time: he related the formal solutions for an
expanding or contracting Universe to redshifts and thus to
observations. Einstein, after having read Friedmann's first paper,
first thought that the solutions are wrong. Later he
admitted that the solutions are mathematically correct, but (in his
opinion) physically
irrelevant. This demonstrates how deeply the idea of a static Universe
was rooted in people's imagination at the time. 

It is often stated that Friedmann was only interested in the
mathematics of the equations, not in their physical content. In our
opinion, this is only partially true. He was certainly strongly
influenced by the mathematicians Weyl and Hilbert, especially the
latter's idea of axiomatization.\footnote{Friedmann had paid a visit
  to G\"ottingen in 1923.} But in his work he strongly emphasized that
the geometry of the world should be determined by theoretical physics
{\em and} observational astronomy.\footnote{In 1924, he even gave a
  thesis topic to his student A.B.~Schechter dealing with the
  question whether trigonometric measurements at astronomical
  dimensions can lead to a decision between different world
  geometries. A paper on this was published three years after
  Friedmann's death by Fr\'edericksz and Schechter [Ellis 1989].}
At the end of his 1922 paper, he gives an estimate of $10^{10}$ years
for the duration of a recollapsing Universe, which is close to the
current estimate for the age of our Universe.

Friedmann was, in particular, interested in the question whether the
world (three-dimensional space) is finite or infinite. This motivated
him to study the 
case of negative curvature in 1924 [Friedmann 1924]. He found that, in
contrast to the spatially closed case discussed in 1922, the case of
negative curvature leaves this question open. He concludes the 1924
paper with the words: ``This is the reason why, according to our
opinion, Einstein's world equations without additional assumptions are
not yet sufficient to draw a conclusion about the finiteness of our
world.''\footnote{The German original reads: ``Dies ist der Grund
  daf\"ur, da\ss, unserer Meinung nach, Einsteins Weltgleichungen
  ohne erg\"anzende Annahmen noch nicht hinreichen, um einen Schlu\ss\
  \"uber die Endlichkeit unserer Welt zu ziehen."} The question
whether it makes sense to talk about actual infinities in physics (in
contrast to mathematics) is an intriguing one and continues to
be discussed up to the present day [Ellis 2018], as
Friedmann's insights continue to inspire modern research.
 \\[1mm]

 \noindent {\bf Friedmann's equations.}\footnote{A comprehensive
    discussion of the material in this and the following
  section can be found in many reviews and textbooks, see
  e.g. [Weinberg~1972], 
  [Mukhanov~2005], [Ellis~2012] and [Calcagni 2017].} 
Starting point is Einstein's field equation
  \be
  \lb{Einstein}
R_{\mu\nu}-\frac12g_{\mu\nu}R+\Lambda g_{\mu\nu}=\frac{8\pi G}{c^2} T_{\mu\nu}.
\ee
Observations indicate that the Universe is approximately isotropic around our
position. These come mainly from the anisotropy spectrum of the Cosmic
Microwave Background (CMB). Adopting the Cosmological Principle (``all
places in the Universe are alike''), one is led to assume
(approximate) isotropy around {\em every} position. One can then
mathematically prove that our Universe must also be (approximately)
homogeneous. The geometry of a homogeneous and isotropic spacetime is
characterised by the line element
\be
\lb{RW}
\D s^2=-c^2\D t^2+a^2(t)\left(\frac{\D r^2}{1-kr^2}+r^2\D\Omega^2\right),
\ee
where $a(t)$ is the scale factor. For the parameter $k$, we have the
possible choices $k=0$ (flat spatial geometry), $k=1$ (positive
curvature), $k=-1$ (negative curvature); only the latter two cases
were treated by Friedmann. Current observations favour 
a spatially flat Universe, although there is still a controversy
[Di~Valentino 2020]. A given value for $k$ does not fix the topology
of our (spatial) Universe, and it is a most intriguing question to
determine the cosmic topology from observations [Luminet~2015]. 

Inserting the ansatz (\ref{RW}) into
(\ref{Einstein}), one is led to Friedmann's equations.\footnote{From
  here on, we set $c=1$.} The first equation is the restriction of the
general Raychaudhuri equation to a homogeneous and isotropic Universe,
\be
\lb{Friedmann1}
\ddot{a}=-\frac{4\pi G}{3}(\rho+3p)a,
\ee
where $\rho$ and $p$ denote energy density and pressure of matter,
respectively. If matter obeys the strong energy condition $\rho+3p\geq
0$, (\ref{Friedmann1}) leads to concave solutions for $a(t)$, that is,
to a world model with a singular origin. The second Friedmann equation
reads
\be
\lb{Friedmann2}
\dot{a}^2=\frac{8\pi G}{3}\rho a^2-k.
\ee
In contrast to (\ref{Friedmann1}), this equation only contains
temporal derivatives up to {\em first} order, so it has the interpretation
of a {\em constraint}. In fact, it is the Friedmann version of the
Hamiltonian constraint in general relativity [Kiefer~2012].

 From
(\ref{Friedmann1}) and (\ref{Friedmann2}), one can derive a third
equation,
\be
\lb{Friedmann3}
\dot{\rho}+3H(\rho+p)=0,
\ee
where $H:=\dot{a}/a$ is the Hubble parameter (its evaluation at the
present day is called Hubble constant, denoted by $H_0$). The
combination $\rho+p$ occuring in (\ref{Friedmann3}) is called inertial mass
density. In these Friedmann equations, we have followed the modern
practice of including the cosmological constant $\Lambda$ into the
density $\rho$ (although this was already suggested by Schr\"odinger
in 1919), because it contributes an `energy density of the vacuum'
$\rho_{\Lambda}:=\Lambda/8\pi G$. Its equation of state 
reads $p_{\Lambda}=-\rho_{\Lambda}$, so from (\ref{Friedmann3}) we see
that $\rho_{\Lambda}$ is constant. For barotropic equations of state
$p=w\rho$, $w\neq-1$, we find from (\ref{Friedmann3}) the solution
\be
\lb{rhoa}
\rho a^{3(1+w)}=\mathrm{constant},
\ee
which includes as particular cases:
\begin{itemize}
\item dust ($p=0$) $\longrightarrow \rho\propto a^{-3}$,
  \item radiation ($p=\rho/3$) $\longrightarrow \rho\propto a^{-4}$,
\item stiff matter ($p=\rho$) $\longrightarrow \rho\propto a^{-6}$.
\end{itemize}
By the kinematic relation $a_0/a=1+z$, with $a_0$ as the present scale
factor, we can relate $\rho$ to the observable redshift $z$ of objects. 
The case of radiation is relevant for the early Universe, while stiff
matter so far seems unrealistic. Today, the Universe is dominated by
dust (about one third) and vacuum energy (about two thirds), leading to the
temporal evolution
\be
\lb{realistic}
a(t)=a_0\left(\frac{3\Omega_0
    H_0}{\Lambda}\right)^{1/3}\sinh^{2/3}\left(\frac32\sqrt{\frac{\Lambda}{3}}t\right),
\ee
where $\Omega_0$ is today's matter density in terms of the critical
density, observationally determined to be about $1/3$. Observations
also indicate that the age of our Universe is about 13.8 billion
years. For large times, the evolution law (\ref{realistic}) asymptotes
to de~Sitter space.\footnote{For late time expansion with constant positive $\rho_\Lambda$ 
 one speaks of {\em dark energy}, but there is also the possibility that
  the effective vacuum energy density varies with time, in which case one 
  speaks of {\em quintessence}. The latter is thought to originate from matter 
  sources and is often modelled by means of a time-dependent scalar field $\phi$.}
 From observations of the CMB, there are strong indications that our
 Universe underwent a quasi-exponential expansion (with very large $\Lambda$)
 already very early
 in its history, a phase called inflation. Inflation offers the means
 to explain the origin of structure in the Universe.\\[1mm]
Instead of barotropic equations of state, one often employs dynamical
matter models, typically with a scalar field $\phi$. In the Friedmann
limit, this field depends, of course, only on time. In the case of a
massless field, it corresponds in (\ref{Friedmann2}) to the choice of
a density $\rho_{\phi}=\dot{\phi}^2/2$.
 
\noindent {\bf Beyond the Friedmann approximation.} 
Beyond the immediate and obvious utility of the Friedmann equations
for cosmological applications there are several important and promising directions  
for future development that build on Friedmann's achievements.
For lack of space we here mention only two of these, namely 
(i) their use for taking first steps towards a theory of quantum gravity, 
and (ii) the generalization of the isotropic ansatz (\ref{RW}) in
order to search for a fundamental symmetry of Nature.

When adapting the ansatz (\ref{RW}) to a quantum mechanical context one speaks
of the so-called {\em minisuperspace approximation}, in which the full
superspace of geometrodynamics, being the moduli space of all
three-metrics modulo spatial diffeomorphisms, is restricted to few
homogeneous degrees of freedom such as the scale factor $a$. This
limit was first discussed by DeWitt in his pioneering paper on
canonical quantum gravity [DeWitt~1967]. 
This is a huge
simplification because key technical issues  
such as the non-renormalizability of perturbative quantum gravity can be ignored
in this approximation. Furthermore, various conceptual issues of
quantum gravity and quantum 
cosmology can be studied. Namely, the direct canonical quantization of
the second  
Friedmann equation (\ref{Friedmann2}) leads to a special case  of the 
Wheeler-DeWitt equation [Kiefer~2012, Calcagni~2017], here given for
the case of a massless homogeneous scalar field,
{\small
\bdm
\left[ \frac{4G\hbar^2}{3\pi a^2} \frac{\partial}{\partial a}\left(
    a\frac{\partial}{\partial a} \right) - 
\frac{\hbar^2}{a^3} \frac{\partial^2}{\partial \phi^2} - \frac{3\pi}{4G}ka 
\right] \Psi(a,\phi) = 0,
\edm}
where a particular factor ordering has been adopted. The wave function 
$\Psi(a,\phi)$ is a simple example of the `wave function of the
universe'. It is, in particular, 
possible to analyse the behaviour of $\Psi$ near the singularity where
$a\rightarrow 0$.   
The Wheeler-DeWitt equation has no external time parameter, but one can
employ the scale 
factor $a$ so as to track the evolution of 
the matter degrees of freedom with respect to this ``intrinsic
time''. (Note that the minisuperspace Wheeler-DeWitt equation is
hyperbolic with respect to $a$.)
Key open issues concern the  
physical interpretation of $\Psi$, the construction of a suitable
Hilbert space, and the meaning of observables;
for a survey and further discussion, see [Kiefer~2012].

The other extension concerns the inclusion of {\em non}-homogeneous
degrees of freedom. 
On the phenomenological side, the evolution of our Universe, if approximated by a  
homogeneous and isotropic spatial part, is successfully described by Friedmann's  
equations, but small inhomogeneities {\em must} be taken into account in
order to understand properties of the CMB in the framework of cosmological 
perturbation theory.  Furthermore, a precise understanding of 
the formation of galaxies and clusters of galaxies requires the numerical treatment 
of the Einstein equations (\ref{Einstein}) and their Newtonian limit.
Incorporating inhomogeneities is likewise crucial for
a better understanding of the origin of the universe, because inhomogeneities 
scale like $a^{-6}$, like stiff matter, and thus dominate very close to 
the Big Bang singularity. This is a crucial issue for inflationary cosmology,
which hinges on the ansatz (\ref{RW}).
Finally, there remain difficult issues related to defining a generally
covariant averaging  
procedure in Einstein's theory that would provide a rigorous basis for the
Friedmann approximation [Buchert 2015].

On the more mathematical side, a key insight came from  the
Belinski-Khalatnikov-Lifshitz (BKL)
analysis [Belinski~1970, 1982] 
of the generic behaviour of solutions of Einstein's equations near a
spacelike singularity. There,
one generalizes the ansatz (\ref{RW}) to
\bdm
\D s^2 = - \D t^2 + a^2(t) \D x^2 + b^2(t) \D y^2 + c^2(t) \D z^2,
\edm
thus giving up isotropy, but retaining spatial homogeneity. A surprising
result to come out of this analysis is the appearance of chaotic
oscillations of the  
metric coefficients $a,b,c$ as one approaches the singularity. This
result indicates 
that the `near singularity limit' of the metric exhibits a far more
complicated behaviour 
than inspection of, say, the Schwarzschild metric would suggest, thus
also showing the 
limitations of the assumption of isotropy.

The BKL analysis has been generalised in many directions, in particular 
also to accommodate inhomogeneities [Krasinski 1997]. Furthermore,
closer study of the BKL limit has revealed evidence for a huge
infinite-dimensional 
symmetry of indefinite Kac-Moody type, vastly generalizing the known
duality symmetries  
of supergravity and string theory. This novel symmetry can possibly
serve as a guiding principle 
towards unifying the fundamental interactions [Damour~2002].

\vskip 2mm We thank Alexander Kamenshchik for his comments on our
manuscript. 
  
\end{multicols}

\newpage

\noindent {\large \bf References}

\vskip 3mm

\noindent [Belinski~1970] V. A. Belinski, I.M. Khalatnikov, I.M. Lifshitz,
\\ \noindent\hspace*{5mm}
Oscillatory approach to a singular point in the relativistic cosmology,
Advances in Physics. 19 (1970) \\ \noindent\hspace*{5mm} 525--573.
\vskip 2mm
\noindent [Belinski~1982] V. A. Belinski, I.M. Khalatnikov, I.M. Lifshitz,
\\ \noindent\hspace*{5mm}
A general solution of the Einstein equations with a time
singularity, 
Advances in Physics 31 (1982) \\ \noindent\hspace*{5mm} 639-667.
\vskip 2mm
\noindent [Buchert~2015] T. Buchert, M.~Carfora,  G.F.R.~Ellis,
E.W.~Kolb, M.A.H.~MacCallum, J.J.~Ostrowski,\\ \noindent\hspace*{5mm}
S.~R\"as\"anen,
B.F.~Roukema, L.~Andersson, A,A,~Coley, D.L.~Wiltshire,
Is there proof that backreaction \\ \noindent\hspace*{5mm} of
inhomogeneities is irrelevant in
cosmology?, Classical and Quantum Gravity, \\ \noindent\hspace*{5mm}
32 (2015) 215021 (38pp). 
\vskip 2mm
\noindent [Calcagni~2017] G.~Calcagni, Classical and Quantum
Cosmology, Graduate Texts in Physics, Springer
\\ \noindent\hspace*{5mm} International
Publishing Switzerland, 2017.
\vskip 2mm
\noindent [Damour~2002] T. Damour, M. Henneaux, H. Nicolai,
\\ \noindent\hspace*{5mm}
$E_{10}$ and a Small Tension Expansion of M Theory, Physical Review
Letters 89 (2002) 221601 (4pp).
\vskip 2mm
\noindent [DeWitt~1967] B.S. DeWitt, Quantum theory of gravity. I.
The canonical theory, Physical Review, \\ \noindent\hspace*{5mm}
       160 (1967)
           1113--1148.
           \vskip 2mm
           \noindent [Di~Valentino~2020] E.~Di~Valentino et al.,
           Cosmology Intertwined~IV: The Age of the Universe and its
           \\ \noindent\hspace*{5mm} Curvature, arXiv:2008.11286v4
           [astro-ph.CO]. 
           \vskip 2mm
           \noindent [Ellis~1989] G.F.R.~Ellis, The Expanding
           Universe: A History of Cosmology from 1917 to 1960,
           in:\\ \noindent\hspace*{5mm} 
           Einstein and the History of General Relativity, ed. by
           D.~Howard and J.~Stachel, Birkh\"auser,
           \\ \noindent\hspace*{5mm} Boston, 1989,~pp.~367--431.
           \vskip 2mm
           \noindent[Ellis~1999] G.F.R. Ellis and A. Krasi\'nski,
           General Relativity and Gravitation 31 (1999) 1985--1990;
           \\ \noindent\hspace*{5mm}
           Addendum: General Relativity and Gravitation {\bf 32}
           (2000) 1937--1938.
           \vskip 2mm
           \noindent [Ellis~2012] G.F.R.~Ellis, R.~Maartens,
           M.A.H.~MacCallum, Relativistic Cosmology, Cambridge
           University  \\ \noindent\hspace*{5mm} Press, Cambridge, 2012.
           \vskip 2mm
           \noindent [Ellis~2018] G.F.R.~Ellis, K.~Meissner,
           H.~Nicolai, The physics of infinity, 
        Nature Physics, \\ \noindent\hspace*{5mm} 14 (2018) 770--772 .
           \vskip 2mm
           \noindent [Friedmann~1922]
           A.~Friedman, \"Uber die Kr\"ummung des Raumes, Zeitschrift
           f\"ur Physik,\\ \noindent\hspace*{5mm} 10 (1922) 377--386.
           \vskip 2mm
           \noindent [Friedmann~1924]
           A.~Friedmann, \"Uber die M\"oglichkeit einer Welt mit
           konstanter negativer Kr\"ummung \\ \noindent\hspace*{5mm} des Raumes,
           Zeitschrift f\"ur Physik, 21 (1924) 326--332.
           \vskip 2mm
           \noindent [Friedmann~2000]
           A.~Friedmann, Die Welt als Raum und Zeit, translated from
           the Russian by G.~Singer, \\ \noindent\hspace*{5mm} Verlag
           Harri Deutsch, Thun und 
           Frankfurt am Main, 2000. The Russian original
           \emph{\foreignlanguage{russian}
            {Мир как пространство и время}}
           appeared 1923 with Academia, Petrograd. In English, the
           title is ``The world as space and 
           time'', but an English translation of the book is not known to us.
           \vskip 2mm
           \noindent [Kiefer~2012] C. Kiefer, Quantum Gravity, 3rd
           ed., Oxford University Press, Oxford, 2012.
           \vskip 2mm
           \noindent[Krasinski~1997]  A. Krasi\'nski,
           Inhomogeneous cosmological models, Cambridge University
           Press, \\ \noindent\hspace*{5mm} Cambridge, 1997.
           \vskip 2mm
           \noindent [Luminet~2015] J.-P.~Luminet, Cosmic Topology,
           Scholarpedia, 10(8):31544. \vskip 2mm
           \noindent [Mukhanov~2005] V. Mukhanov, Physical Foundations of Cosmology,
                 Cambridge University Press (2005). \vskip 2mm
          \noindent [Weinberg~1972] S. Weinberg, Gravitation and Cosmology, 
          John Wiley and Sons (1972).
           

\end{document}